\documentclass[%
 reprint,
 amsmath,amssymb,
 aps,
]{revtex4-2}
\usepackage[colorlinks=true,urlcolor=blue,linkcolor=blue,citecolor=blue]{hyperref}
\usepackage{amsmath} 
\usepackage{amssymb}
\usepackage{epsf}
\usepackage{color}
\usepackage{dcolumn}
\usepackage{bm}

\usepackage{graphicx}
\graphicspath{ {./images/} }

\begin{document}

\title{A new microscopic representation of the spin dynamics in quantum systems with biquadratic exchange interactions}%

\author{Mariya Iv. Trukhanova$^{1,2}$}
\email{trukhanova@physics.msu.ru}
\author{Pavel Andreev$^{1}$}
\email{andreevpa@physics.msu.ru}

\affiliation{$^{1}$~Faculty of Physics,
Lomonosov Moscow State University, Leninskie Gory-1, 119991 Moscow, Russia  \vspace{1em} \\
$^{2}$~Theoretical Physics Laboratory, Nuclear Safety Institute, Russian Academy of Sciences,
B. Tulskaya 52, 115191 Moscow, Russia}

\begin{abstract}
  There is a version of the Landau-Lifshitz equation that takes into account the Coulomb exchange interactions between atoms, expressed by the term $\sim\bm{s}\times\triangle\bm{s}$. On the other hand, ions in the magnetic materials have several valence electrons on the $d$-shell, and therefore the Hamiltonian of many-electron atoms with spins $S>1$ should include a biquadratic exchange interaction. We first propose a new fundamental microscopic derivation of the spin density evolution equation with an explicit form of biquadratic exchange interaction using the method of many-particle quantum hydrodynamics. The equation for the evolution of the spin density is obtained from the many-particle Schrodinger-Pauli equation and contains the contributions of the usual Coulomb exchange interaction and the biquadratic exchange. Furthermore, the derived biquadratic exchange torque in the spin density evolution equation is proportional to the nematic tensor for the medium of atoms with spin $\textit{S = 1}$. Our method may be very attractive for further studies of the magnetoelectric effect in multiferroics.
\end{abstract}

\maketitle

\section{Introduction}
Interactions between the identical particles lead to the spin dependence of their energy \cite{Stohr}. The quantum mechanical nature of the Coulomb exchange interactions was demonstrated by Heisenberg \cite{Heisenberg}, \cite{Lvov}. The description of ferromagnetic and antiferromagnetic ordering is in good agreement with the model based on the effective spin Hamiltonian of the exchange interactions of atoms (ions)  \cite{Lvov}
\begin{equation} H= J^{(1)}(\bm{r}_{ij})\bm{S}_i\cdot\bm{S}_j+J^{(2)}(\bm{r}_{ij})(\bm{S}_i\cdot\bm{S}_j)^2,\label{H}\end{equation}where $J^{(1)}$, $J^{(2)}$ are the exchange integrals, and $\bm{S}_i, \bm{S}_j$ are the spins in the nodes of the lattice. The first term is an antisymmetric exchange interaction and corresponds to the  \textit{Heisenberg Hamiltonian}. The second term, called the biquadratic exchange, needs to be considered for many-electron ions with spins $S >1/2$, and vanishes in the lowest order in the overlap parameter of the ions wave functions. The Ising model, see \cite{Tm}, was developed for two- and three-dimensional spin lattices at low temperatures.

 The description of the magnetoelectric effect in multiferroic materials, is based on exchange interactions \cite{Mm, Landau}. The microscopic explanation of the ferroelectricity of spin origin is based on the following mechanisms: symmetric exchange interactions between the neighboring ions \cite{Tokura}, antisymmetric exchange interactions between noncollinear spins \cite{Sergienko, Nagaosa}, and spin-dependent metal-ligand hybridization \cite{Arima}. The symmetric exchange interactions lead to the striction along a crystallographic direction $\bm{\Pi}_{ij}$ \cite{Tokura}, so that the polarization appears as a result of the inversion symmetry breaking: $\bm{P}_{ij} \sim \bm{\Pi}_{ij}(\bm{S}_i\cdot\bm{S}_j)$. In the case of antisymmetric exchange interactions, the polarization is induced under the influence of spin-orbital coupling \cite{D}. The relativistic corrections to Anderson’s superexchange between magnetic ions in the presence of spin-orbital coupling lead to an antisymmetric exchange, also named as Dzyaloshinskii-Moriya interaction \cite{D, Moria}, with \textit{Hamiltonian} $H_{DM}=\bm{D}_{ij}\cdot\bm{S}_i\times\bm{S}_j$, where $\bm{D}_{ij}$ is the Dzyaloshinskii vector. The polarization that appears due to this mechanism is determined by the expression: $\bm{P}_{ij} \sim \bm{e}_{ij}\times(\bm{S}_i\times\bm{S}_j)$, where $\bm{e}_{ij}$ is a unit vector connecting two spins \cite{Sergienko}.

A phenomenological theory of inhomogeneous ferroelectric magnets was proposed by M. Mostovoy and describes their thermodynamics and the behavior of polarization in the magnetic field \cite{Mostovoy}. In this research, the Ginzburg-Landau approach was applied to derive the expression for the electric polarization induced by spiral spin-density wave states. A few years later, Masahito Mochizuki suggested a microscopic model of the multiferroic perovskite manganites \cite{Masahito}. In this model, the calculation is carried out for each individual ion at the site of the crystal lattice. The Landau-Lifshitz-Gilbert equation was derived for the Mn $3d$-spin atoms in the local magnetic field from the Heisenberg model with additional interactions and magnetic anisotropies. 

The theory of the ferroelectric polarization which is induced by the spin-spiral order was developed in Refs.\cite{Xiang1, Xiang2}. In particular, a method for describing the electronic and ion-displacement contributions to the spin-order-induced ferroelectric polarization was proposed in Ref. \cite{Xiang2}. For the evolution of the electric polarization with time, the nonlinear Duffing oscillator equation was obtained using the Euler-Lagrange equation \cite{Band}. A phenomenological framework for studying the switching dynamics of coupled polarization and magnetization in single-phase multiferroic materials was developed in Ref. \cite{Kuntal}, where the oscillatory mode of magnetization can act as a microwave signals.

Understanding the microscopic foundations of the magnetoelectric effect in magnetically ordered media is an important step in the development of polarization control \cite{Choi}. The magnetically induced electric dipole moment of the cell of the crystal lattice, the collective dynamics of these dipoles, and the formation of wave patterns are realized in three-dimensional physical space. The collective effects require representation of the dynamics in terms of field variables defined in three-dimensional physical space \cite{Andreev01}. A model that describes the dynamics of non-equilibrium systems can be developed using the method of many-particle quantum hydrodynamics \cite{Andreev} - \cite{Andreev4}. 

Exchange interactions underlie spin ordering in magnetic structures. Magnetocrystalline anisotropy, external magnetic field, biquadratic exchange, and demagnetization field can also influence the formation of magnetic structures. The basic dynamical equation for magnetization was derived in 1935 by Landau and Lifshitz \cite{Landau2}. Later, using the Lagrangian approach, Hilbert supplemented the damping term with this equation \cite{Gilbert2, Stiles, Lak}. 
The exchange interaction is electrostatic in nature and is directly related to the Pauli principle. While the symmetric Heisenberg exchange interaction is well studied and its contribution to the Landau-Lifshitz equation has long been derived \cite{Lvov, Atxitia}, biquadratic exchange is rarely seen in theoretical articles regarding magnetic phenomena. At the same time, magnetic atoms and ions have several valence electrons on the $d$-shell, and therefore the Hamiltonian of many-electron atoms has to include a biquadratic exchange. Since most of the magnetic ions in type-II multiferroics have spin $S > 1/2$ \cite{Nagaosa, Tokura}, it is necessary to consider the biquadratic exchange interaction in the equation of magnetization evolution (spin density). 

In this article, we propose a fundamental microscopic derivation of the magnetization evolution equation, or the Landau-Lifshitz-like equation with biquadratic exchange interaction. As a starting point, the many-particle Schrodinger-Pauli equation for a system of particles (ions or atoms) is used. Further, the exchange interactions between individual electrons are considered in the form of the Heisenberg-like Hamiltonian of interactions of the magnetic moments of ions. We derive the spin density evolution equation, which includes the contribution of the Coulomb exchange interaction between valence electrons and the biquadratic exchange interactions. For this purpose, we use the method of many-particle quantum hydrodynamics.

\section{The theoretical model}
Ferromagnetic materials have been studied using different theoretical methods for a long time. In this article, we propose a new approach to learning collective phenomena in magnetically ordered materials. The many-particle quantum hydrodynamics method has been developed for several physical systems, such as quantum plasmas \cite{Andreev5}, polarized Bose-Einstein condensates \cite{Andreev6} and spin-separated quantum systems \cite{Andreev7}. The first step in our approach is the introduction of the many-particle interaction Hamiltonian of interactions. The second step is to determine the microscopic representation of the spin density of the particle system. In order to obtain the equation defining the dynamics of spin density, it is imperative to identify the macroscopic form of the spin current density and the torque density.

We consider a system of magnetic ions localized in the nodes of the crystal lattice and denote by $\bm{S}_j$ the spins of the valence electrons of ions. We construct the model for two kinds of ions: those with spin $S=1/2$ and $S=1$. But, this model can also approximately describe the behavior of spins $S>1/2$. Most of the magnetic ions in ferroelectric materials have spins greater than one. For instance, the perovskite oxide with an orthorhombic structure, TbMnO$_3$, has magnetic ions Mn$^{3+}$ with $S$=2, or the cubic spinel ZnCr$_2$Se$_4$, where Cr$^{3+}$ has spins $S$=$3/2$, or the mixed-valence manganese oxide family, RMn$_2$O$_5$, R is Y, Tb, Ho, Er, Tm, with Mn$^{4+}$ ($S$=3/2). This means that our theoretical model has to include higher-order exchange interactions.

The first step of our approach is the introduction of the many-particle interaction Hamiltonian in the form
$$ \hat{H}=\sum^N_{j=1}\left(\frac{\hat{\bm{p}}_j^2}{2m_j}-\gamma_j\hat{S}^{\beta}_jB^{\beta}_{j}\right)$$\begin{equation}\label{H0}+\frac{1}{2}\sum_{l\neq j}^N\left(U^{(1)} (\bm{r}_{lj})\hat{S}^{\beta}_l\hat{S}^{\beta}_j+U^{(2)}(\bm{r}_{lj})(\hat{S}^{\beta}_l\hat{S}^{\beta}_j)^2\right),\end{equation}
where $\hat{p}^{\alpha}_j=-i\hbar\partial^{\alpha}_j$ is the momentum operator,  $m_j$, $\gamma_j$ are the mass and  gyromagnetic ratio of $j$-th particle. The first term in the above Hamiltonian (\ref{H0}) is the sum of the kinetic energies of all particles. The second term is responsible for the action of the external magnetic field $\bm{B}_{j}$ on the particle's magnetic moment. The third and fourth terms characterize Coulomb (symmetric and biquadratic) exchange interactions. In real magnetic materials and multiferroics, the interactions described above occur between two neighbouring spins of magnetic ions located at the nodes of the crystal lattice. For instance, the positively charged transition metal ions Ni$^{2+}$ in the case of Ni$_3$V$_2$O$_8$. In the simplest case of one-electron ions, the exchange interaction is proportional to the scalar product of spins. On the other hand, the quadratic correction is to be taken into consideration for many-electron systems and for the spin-ordering of low-dimensional nanostructures. The biquadratic exchange determines the magnetic features in multilayer materials \cite{Sl}, perovskites and some oxides \cite{Fedorova, Kartsev}, and are important for the magnetic properties of 2D materials \cite{Kartsev}.

The generalized Hamiltonian (\ref{H0}) contains effective potentials $U^{(1)}(\bm{r}_{lj})$ and $U^{(2)}(\bm{r}_{lj})$, which are functions of the distance between the interacting particles $\bm{r}_{lj}$. These effective potentials are the exchange integrals. Positive (or negative) values of the function $U^{(1)}(\bm{r}_{lj})$ characterize the antiferromagnetic or ferromagnetic equilibrium states. The dipole-dipole interactions between magnetic moments are not considered in the framework of the many-particle Hamiltonian (\ref{H0}). In this article, we aim to investigate Coulomb exchange interactions. But, the method of many-particle quantum hydrodynamics was used to investigate physical systems with dipole-dipole magnetic interactions (see Refs. \cite{Andreev8, Andreev9}).

Collective processes in many-particle systems are described by macroscopic physical fields. Let's study the collective dynamics using the local spin density of particles as a suitable collective variable. For this purpose the dynamics of $N$-interacting particles is given by the Pauli-Schrodinger equation $i\hbar\partial_t\psi_s(R,t)=\hat{H}\psi_s(R,t)$ with Hamiltonian (\ref{H0}). In the approach of many-particle quantum hydrodynamics, the many-particle wave function $\psi_s(R,t)$ depends on the $3N$ position variables $R=\{\bm{r}_1,...\bm{r}_N\}$ and time. The spin density in the neighborhood of $\bm{r}$ can be introduced as the quantum average of the spin density operator. The average value of the operator of the physical quantity coincides with the eigenvalue of the operator if the quantum state is an eigenstate of the operator of the physical quantity \cite{Andreev3}. The operator of spin density is chosen as the quantization of the microscopic local spin density, and is given by $\hat{\bm{s}}^{\alpha}=\sum_{k}\delta(\bm{r}-\bm{r}_k)\hat{\bm{S}}_k^{\alpha}$. As a result, the spin density can be introduced in the form  $$\bm{s}^{\alpha}(\bm{r},t)=\sum_{s=s_1,...,s_N}\int dR\sum_{k=1}^N\delta(\bm{r}-\bm{r}_k)\qquad\qquad$$\begin{equation}\label{Spin}\qquad\qquad\qquad\qquad\times\psi^{\dagger}_{s}(R,t)\hat{\bm{S}}_k^{\alpha}\psi_{s}(R,t), \end{equation}where $dR=\Pi_{j=1}^Nd\bm{r}_j$ is the element of the volume in $3N$-dimensional configuration space, $N$ is the number of particles, and $\hat{\bm{S}}_k^{\alpha}$ is the spin operator of the $k$-th particle.

For the case of $S = 1/2$, the form of the cartesian components of the spin operator in the cyclic basis is $\hat{\bm{S}}_k=\frac{1}{2}\hat{\bm{\sigma}}_k$, where the Pauli matrices are given by
\begin{equation}\qquad\hat{\sigma}^x_k= \begin{pmatrix} 0 & 1 \\ 1 & 0
\end{pmatrix},  \hat{\sigma}^y_k= \begin{pmatrix} 0 & -i \\ i & 0 \end{pmatrix},   \hat{\sigma}^z_k=\begin{pmatrix}
 1 & 0 \\ 0 & -1 \end{pmatrix}.\end{equation}
 In the case of $S=1$, the spin operator can be represented in the form $\hat{\bm{S}}_k=\hat{\bm{F}}_k$, where the spin matrices are chosen as
$$
\hat{F}^x_k=\frac{1}{\sqrt{2}} \begin{pmatrix} 0 & 1 & 0 \\ 1 & 0 & 1 \\ 0 & 1 & 0 \end{pmatrix},$$ \begin{equation}\hat{F}^y_k= \frac{1}{\sqrt{2}}\begin{pmatrix} 0 & -i & 0 \\ i & 0 & -i \\ 0&i&0 \end{pmatrix}, \;  \hat{F}^z_k= \begin{pmatrix} 1 & 0 & 0 \\ 0 & 0 & 0 \\ 0 & 0 & -1 \end{pmatrix}   \end{equation}
In the next step, we derive the dynamical equation for the evolution of the spin density. For this purpose, we take the time derivative of the spin density (\ref{Spin}). This derivative acts on the many-particle wave function $\psi_s(R,t)$ under the integral. The value of $\dot{\psi}_s$ is determined by the Pauli-Schrodinger equation with Hamiltonian (\ref{H0}). The resulting equation for the evolution of the spin density is derived in the form$$\partial_t{s}^{\alpha}(\bm{r},t)+\partial_{\beta}j^{\alpha\beta}_s(\bm{r},t)\qquad\qquad\qquad$$\begin{equation}\label{S1}\qquad\qquad=\frac{2\gamma}{\hbar}\epsilon^{\alpha\beta\mu}s^{\beta}(\bm{r},t)B^{\mu}(\bm{r},t)+T^{\alpha}(\bm{r},t).\end{equation}Let us now discuss the relation between the equation (\ref{S1}) and the usual Landau-Lifshitz equation ($LL$). The $LL$ equation describes the evolution of the magnetization of magnetically ordered materials in the effective magnetic field and leads to many physical effects \cite{Bogdan}. In particular, the precession of the magnetic moments in the external magnetic field is the simplest consequence of this equation. This effect is characterized by the first term on the right side of Eq. (\ref{S1}). The precession of the magnetic moments can be caused by particle interactions, for instance, dipole-dipole interactions, which were investigated in Refs. \cite{Andreev8, Andreev9}. The Coulomb exchange interaction considered in this paper is taken into account by the last term on the right side of Eq. (\ref{S1}). This effect is also called the energy of heterogeneity in phenomenological consideration. On the other hand, the Landau-Lifshitz-Gilbert ($LLG$) equation takes into account the damping term \cite{Lvov}. But, in this article, we focus on the microscopic mechanisms of the Coulomb exchange interactions for $S>1/2$ systems and do not include any damping because there are no degrees of freedom that can be used as an energy reservoir. 

The second term on the left-hand side of Eq. (\ref{S1}) represents the spatial derivative of the spin current density. The spin current tensor is a product of two vectors: the charge current density and spin polarization \cite{M}. The spin-polarized current of electrons determines the spin transfer torque and leads to the additional precession of the magnetization \cite{Nan}. In other words, if the spin current $j^{\alpha\beta}_s$ changes from point to point, the magnetization or the spin density vector experiences spin transfer torque.
The microscopic representation of the spin current density can be derived using the many-particle quantum hydrodynamics method in the form $$j^{\alpha\beta}_s(\bm{r},t)=\sum_{s=s_1,...,s_N}\int dR\sum_{k=1}^N\delta(\bm{r}-\bm{r}_k)\frac{1}{2m_k}$$\begin{equation}\times\left((\hat{p}^{\beta}_k\psi_s)^{\dagger}(R,t)\hat{\bm{S}}^{\alpha}_k\psi_s(R,t)+h.c.\right).\end{equation} The above expression is not postulated. Details of the derivation of the spin current density can be found in Refs. \cite{Andreev}, \cite{Andreev8}, \cite{Andreev9}.
\subsection{Biquadratic exchange interactions}
Let's discuss Eq. (\ref{S1}) in details. The first term on the right-hand side of the equation is responsible for the external magnetic field $\bm{B}$. The second term characterizes the torque acting on the spin density via the Coulomb exchange interactions $T^{\alpha}(\bm{r},t)$ and includes potentials $U^{(1)}_{lj}$ and $U^{(2)}_{lj}$. After the calculations, this torque can be represented in the following form $$T^{\alpha}(\bm{r},t)=\frac{1}{2\hbar}\epsilon^{\alpha\beta\gamma}\sum_s\int dR\sum_{l\neq j}^N\biggl(\delta(\bm{r}-\bm{r}_j)-\delta(\bm{r}-\bm{r}_l)\biggr)$$
\begin{equation}\label{T1} \psi^{\dagger}_s\left(U^{(1)}_{lj}\hat{S}^{\beta}_l\hat{S}^{\gamma}_j+\frac{1}{2}U^{(2)}_{lj}\left(4\hat{\pi}^{\beta\mu}_l\hat{\pi}^{\gamma\mu}_j-\hat{S}^{\beta}_l\hat{S}^{\gamma}_j\right)\right)\psi_s,\end{equation} where the nematic tensor operator is derived in the form \begin{equation}\hat{\pi}^{\gamma\mu}_l=\frac{1}{2}\biggl(\hat{S}^{\gamma}_l\hat{S}^{\mu}_l+\hat{S}^{\mu}_l\hat{S}^{\gamma}_l\biggr),\end{equation}here we use the expression $[\hat{S}^{\alpha}_l,\hat{S}^{\beta}_j]=i\epsilon^{\alpha\beta\gamma}\delta_{lj}\hat{S}^{\gamma}_l.$ The higher spin systems support spin-nematic order in addition to the charge and magnetic ones. The magnetic and nematic orders compete because increasing one requires lowering the other \cite{Ueda}.

Let us derive a more explicit form of the torque vector $\bm{T}$.
We consider the identical particles and introduce variables of the center of the mass for each pair of particles $\bm{R}_{lj}=\frac{1}{2}(\bm{r}_l+\bm{r}_j)$ and the relative distance between them $\bm{r}_{lj}=\bm{r}_l-\bm{r}_j$. The potentials $U^{(1)}_{lj}(\bm{r}_{lj})$ and $U^{(2)}_{lj}(\bm{r}_{lj})$ decrease rapidly as the relative distance between particles increases $r_{ij}$. The delta-functions $\delta(\bm{r}-\bm{r}_j)$, $\delta(\bm{r}-\bm{r}_l)$, the $l$-th and $j$-th arguments of the wave function $\psi_s(...,\bf{r}_l,...,\bf{r}_j,...,t)$ are subjected to a Taylor series expansion. For this purpose, the many-particle wave function can be represented in the form
$\psi_s(R,t)=\psi_s(\bm{r}_1,...,\bm{R}_{lj}+\frac{1}{2}\bm{r}_{lj},...,\bm{R}_{lj}-\frac{1}{2}\bm{r}_{lj},...,\bm{r}_N,t),$ and can be decomposed into a Taylor series by a small parameter $\bm{r}_{lj}$
  $$\psi_s(R,t)=\psi_s(R',t)+\frac{1}{2}r^{\alpha}_{lj}\partial^{\alpha}_{R,1}\psi_s(R',t)\qquad\qquad$$ \begin{equation}\qquad\qquad\qquad\qquad-\frac{1}{2}r^{\alpha}_{lj}\partial^{\alpha}_{R,2}\psi_s(R',t),
\end{equation}where $R'=\bm{r}_1,...,\bm{R}_{lj},...,\bm{R}_{lj},...,\bm{r}_N$, $\partial_{R,1}^{\alpha}=\frac{\partial}{\partial\bm{R}_{lj}}$ is the partial derivative with respect to $\bm{R}_{lj}$ at the $l$-th place and  $\partial_{R,2}^{\alpha}=\frac{\partial}{\partial\bm{R}_{lj}}$ is the partial derivative with respect to $\bm{R}_{lj}$ at the $j$-th place. Integral over $\bm{r}_{lj}$ and integral over $\bm{R}_{lj}$ are independent, so that we can separate them explicitly $dR=dR_{N-2}d\bm{R}_{lj}d\bm{r}_{lj}$. 
As a result, in the limit of the small relative distance between particles $\bm{r}_{lj}$, in the second order of decomposition and after the serious of calculations, we derive the non-trivial solution for the exchange torque
$$
  T^{\alpha}(\bm{r},t)=(\hat{g}^{(1)}+\hat{g}^{(2)})\epsilon^{\alpha\beta\gamma}s^{\beta}(\bm{r},t)\triangle s^{\gamma}(\bm{r},t)\qquad\qquad
$$\begin{equation}\label{T4}
  \qquad\qquad\qquad\qquad-4\hat{g}^{(2)} \epsilon^{\alpha\beta\gamma}\pi^{\beta\mu}(\bm{r},t)\triangle \pi^{\gamma\mu}(\bm{r},t),
\end{equation}where the constants of interactions have the form $$\hat{g}^{(1)}=-\frac{1}{6\hbar}\int \xi^2U^{(1)}(\xi)d\bm{\xi}, \qquad$$ 
\begin{equation}\qquad\qquad\hat{g}^{(2)}=\frac{1}{12\hbar}\int \xi^2 U^{(2)}(\xi)d\bm{\xi}. \end{equation}  

Let's discuss the physical meaning of terms in the definition \eqref{T4}. For $U^{(2)}=0$, the first term in the expression (\ref{T4}) is proportional to $\hat{g}^{(1)}$ and follows from the usual Coulomb exchange interaction (symmetric exchange) \cite{Ivanov}. While the equation for the spin density evolution (\ref{S1}) in the absence of the external magnetic field and for the case of $U^{(2)}=0$ has the form of the continuity equation
\begin{equation}
\partial_t\bm{s}^{\alpha}(\bm{r},t)+\partial_{\mu}\cdot\bm{j}^{\alpha\mu}_{exc}(\bm{r},t)=0,
\end{equation}where the expression $j^{\alpha\mu}_{exc}(\bm{r},t)=-\hat{g}^{(1)}\epsilon^{\alpha\beta\gamma}s^{\beta}(\bm{r},t)\partial^{\mu}s^{\gamma}(\bm{r},t)$ represents the exchange spin current.  The above equation is the spin angular-momentum conservation law and has been derived in Refs. \cite{Lvov, Atxitia} and sometimes can be represented through the exchange magnetic field $\partial_t\bm{s}=\gamma\bm{s}\times\bm{H}_{exc}$, where $\bm{H}_{exc}=A_{exc}\triangle\bm{s}$ denotes the exchange field, $A_{exc}$ is a material constant \cite{Cimrák}.

The presence of the biquadratic exchange interaction, which is taken into account by potential $U^{(2)}(\xi)$, makes the situation different. Although the higher-order exchange interactions have already been studied in the literature \cite{Kartsev, Brinker, Mila}, the equation for the evolution of the spin density in the presence of the biquadratic exchange interactions has been obtained in this article for the first time. The biquadratic exchange leads to the non-vanishing exchange spin current
$$\bm{j}_{exc}^{'\alpha\mu}(\bm{r},t)=-\hat{g}^{(2)}\epsilon^{\alpha\beta\gamma}s^{\beta}(\bm{r},t)\partial^{\mu}s^{\gamma}(\bm{r},t)\qquad\qquad$$
 \begin{equation}\qquad\qquad\qquad+4\hat{g}^{(2)}\epsilon^{\alpha\beta\gamma}\pi^{\beta\mu}(\bm{r},t)\partial^{\mu} \pi^{\gamma\mu}(\bm{r},t),\end{equation} where the density of nematic tensor is derived in the microscopic form $$\pi^{\beta\mu}(\bm{r},t)=\sum_{s}\int dR\sum_{k=1}^N\delta(\bm{r}-\bm{r}_k)\qquad\qquad$$\begin{equation}\label{nematic}\qquad\qquad\qquad\qquad\times\psi^{\dagger}_{s}(R,t)\hat{\pi}^{\beta\mu}_k\psi_{s}(R,t).\end{equation} For ions with spin-1/2, the density of the nematic tensor becomes proportional to the concentration $\pi^{\alpha\beta}=(1/4)\delta^{\alpha\beta}n(\bm{r},t)$.
\subsection{The discussion of the results}
The Landau-Lifshitz equation was first derived from phenomenological considerations in 1935 and provides the basic equation for the evolution of magnetization in bulk magnetics \cite{Landau2}. The effect of relativistic interactions, the damping term, was introduced at a later date by Gilbert, and the conclusive equation was called the Landau-Lifshitz-Gilbert equation \cite{Gilbert2}. This equation allows us to take into account all possible fundamental interactions, first of all the influence of the external magnetic field, but also the Coulomb exchange interaction, which is a consequence of the overlap of the wave functions of electrons \cite{Stiles}. The magnetocrystalline anisotropy field, the demagnetization field, and biquadratic exchange interactions can also be included into the equation. We focus our attention specifically on the biquadratic exchange interaction, since there is no clear derivation of its effect on the evolution of magnetization. In magnetic atoms and metal ions, the $d$-shell is partially filled. Therefore, several valence electrons of the atom, for which the values of spins will be $S > 1/2$, participate in the Coulomb exchange interaction. This means that for most many-electron magnetic atoms, the Hamiltonian of the exchange interaction between two atoms must be different from the \textit{Heisenberg Hamiltonian}. However, in the first approximation, in most studies, the exchange of only one pair of electrons is taken into account by the overlapping wave functions of magnetic atoms. As a result, the authors also use the Heisenberg approximation for many-electron atoms, but, generally, it is not correct. For magnetic atoms with $S>1/2$, such an approximation is often insufficient, and it is necessary to take into account the biquadratic exchange, which we have included in the Hamiltonian of the system \eqref{H}. In the energy of magnetic inhomogeneity, an additional term also appears and corresponds to biquadratic exchange. 

On the one hand, the exchange interaction between valence electrons of many-electron metal atoms has been considered in a large number of articles and has a long history \cite{Lvov}, but to obtain the influence of this exchange on the evolution of magnetization is a difficult task. In this research, using the method of many-particle quantum hydrodynamics, we first obtained an explicit form of the biquadratic exchange interaction in the Landau-Lifshitz-like equation. We derived the effect of this interaction from the basis of the theoretical model and introduced a microscopic representation of the nematic tensor \eqref{nematic} that arises in the model as a result of the derivation of the equation.

\section{Conclusions}
 In this article, we present a new fundamental microscopic derivation of the spin density evolution equation (\ref{S1}) with torque (\ref{T4}) for a magnetically ordered material. The equation (\ref{S1}) is derived as part of the system of equations for the evolution of the macroscopic characteristics of the medium. Although we consider the magnetic ions in the nodes of the crystal lattice, the structure of the equation (\ref{S1}) is analogous to the equation of hydrodynamics. We aim to study the influence of Coulomb exchange interactions on spin density dynamics. We focus on the microscopic derivation of the exchange interactions using the method of many-particle quantum hydrodynamics. As a result of derivation, the biquadratic exchange interaction accounting leads to the torque in the equation for the evolution of the spin density.

The first step in describing the magnetoelectric effect in multiferroics is a new model that takes into account the effect of exchange interactions on the dynamics of spin density. Explicitly accounting for the impact of biquadratic exchange is the first major step in this direction. Many-particle quantum hydrodynamics allows us to study the dynamics of magnetization and polarization in systems with exchange interactions. Moreover, this method allows us, from first principles, to take into account all the necessary interactions between particles and obtain microscopic definitions of the main macroscopic observables, such as spin density, magnetization, and electric polarization. This unique feature makes this method very attractive for further studies of the magnetoelectric effect in multiferroics.


\begin{acknowledgments}
The research of Mariya Iv. Trukhanova is supported by the Russian Science Foundation under the grant No. 22-72-00036 (https://rscf.ru/en/project/22-72-00036/).
\end{acknowledgments}


\end{document}